\documentclass[preprint,nofootinbib,floats,superscriptaddress,tightenlines,amsfonts,amsmath,amssymb]{revtex4}
\pdfoutput=1
\usepackage{graphicx}
\usepackage{bm}
\usepackage{color}
\usepackage{hyperref}
\usepackage{multirow}
\usepackage{slashed}

\newcommand{\calO}{{\cal O}}
\newcommand{\GeV}{\rm GeV}

\newcommand{\Hc}{\rm H. c. }

\begin{document}
\baselineskip=17pt \parskip=5pt

\preprint{NCTS-PH/2003}
\hspace*{\fill}

\title{Breaking the Grossman-Nir Bound in Kaon Decays}

\author{Xiao-Gang He}\email{hexg@phys.ntu.edu.tw}

\affiliation{Department of Physics, National Taiwan University,\\
No.\,\,1, Sec.\,\,4, Roosevelt Rd., Taipei 106, Taiwan}
\affiliation{Physics Division, National Center for Theoretical Sciences,\\
No.\,\,101, Sec.\,\,2, Kuang Fu Rd., Hsinchu 300, Taiwan}

\author{Xiao-Dong Ma}\email{maxid@phys.ntu.edu.tw}
\affiliation{Department of Physics, National Taiwan University,\\
No.\,\,1, Sec.\,\,4, Roosevelt Rd., Taipei 106, Taiwan}

\author{Jusak Tandean}\email{jtandean@phys.ntu.edu.tw}
\affiliation{Department of Physics, National Taiwan University,\\
No.\,\,1, Sec.\,\,4, Roosevelt Rd., Taipei 106, Taiwan}
\affiliation{Physics Division, National Center for Theoretical Sciences,\\
No.\,\,101, Sec.\,\,2, Kuang Fu Rd., Hsinchu 300, Taiwan}

\author{German Valencia}\email{german.valencia@monash.edu}
\affiliation{School of Physics and Astronomy, Monash University, Melbourne VIC-3800, Australia
\bigskip}


\begin{abstract}
The ratio $\mathcal{B}(K_L\to\pi^0\nu\bar\nu)/\mathcal{B}(K^+\to\pi^+\nu\bar\nu)$ of the branching fractions of kaon decays $K_L\to\pi^0\nu\bar\nu$ and $K^+\to\pi^+\nu\bar\nu$ has a maximum of about 4.3 under the assumption that the underlying interactions change isospin by $\Delta I=1/2$.
This is referred to as the Grossman-Nir (GN) bound, which is respected by the standard model (SM) and by many scenarios beyond it.
Recent preliminary results of the KOTO and NA62 Collaborations searching for these kaon modes seem to imply a violation of this bound.
The KOTO findings also suggest that $\mathcal{B}(K_L\to\pi^0\nu\bar\nu)$ could be much larger, by nearly two orders of magnitude, than that predicted in the SM. In this work we study the possibility of violating the GN bound in an effective field theory approach with only SM fields. 
We show that the bound holds, in addition to the original GN scenarios,  whether or not the kaon decays conserve lepton number. 
We demonstrate that the inclusion of $\Delta I=3/2$ operators can lead to a violation of the GN bound and illustrate with an example of how the KOTO numbers may be reached with a new physics scale of order tens of GeV.

\end{abstract}

\maketitle

\section{Introduction}

Recently the KOTO Collaboration  has presented a preliminary report on its latest search
for the flavor-changing neutral current (FCNC) decay of neutral kaon $K_L$ into a neutral pion $\pi^0$ and an unobserved neutrino pair (2$\nu$), having achieved a single event sensitivity of $6.9\times 10^{-10}$ and showing 3 candidate events in the signal region~\cite{koto}. Should these turn out to be real signal events, they would imply a branching fraction of ${\mathcal B}(K_L\to\pi^02\nu)\simeq2.1\times 10^{-9}$, which is greater by almost two orders of magnitude than the standard model (SM) prediction~\cite{Charles:2004jd,ckmfitter,Tanabashi:2018oca}: ${\mathcal B}(K_L\to\pi^0\nu\bar\nu)_{\rm SM} = (3.0\pm0.2)\times 10^{-11}$.
If confirmed in the future, this huge enhancement seen by KOTO would constitute early evidence for new physics (NP) beyond the SM in the quark sector. For comparison, KOTO~\cite{Ahn:2018mvc} earlier set the limit ${\mathcal B}(K_L^0\to\pi^02\nu)_{\rm KOTO15}<3.0\times10^{-9}$ at 90\% confidence level (CL).

This process is related to its charged counterpart $K^+\to\pi^+2\nu$, which has a SM branching fraction of~\cite{Charles:2004jd,ckmfitter,Tanabashi:2018oca} ${\mathcal B}(K^+\to\pi^+\nu\bar\nu)_{\rm SM} = (8.5\pm0.5)\times 10^{-11}$.
The latest quest for it by the NA62 Collaboration~\cite{na62} has yielded a preliminary bound of
${\mathcal B}(K^+\to\pi^+2\nu) < 1.85\times 10^{-10}$ at 90\%\,\,CL.
This complements the previous finding of BNL E787/949~\cite{Artamonov:2008qb,Artamonov:2009sz}: 
${\mathcal B}(K^+\to\pi^+\nu\bar\nu)_{\rm E949}=\big(1.73_{-1.05}^{+1.15}\big) \times 10^{-10}$.

Evidently, the aforesaid new information from KOTO and NA62 suggests the possibility that the experimental value of the ratio $r_{\cal B} = {\cal B}(K_L\to\pi^02\nu)/{\cal B}(K^+\to\pi^+2\nu)$ exceeds the so-called Grossman-Nir (GN) upper bound~\cite{Grossman:1997sk}, which is about 4.3.
It goes without saying that a clear violation of this bound would have important consequences. In this paper we investigate the possibility of such a violation within a scenario with no new particles at the GeV scale and with or without lepton number and lepton flavor conservations.\footnote{Several different scenarios involving new particles contributing to $K \to \pi$ plus missing energy to break the GN bound have been studied recently~\cite{Fuyuto:2014cya,Kitahara:2019lws,Fabbrichesi:2019bmo, Egana-Ugrinovic:2019wzj, Dev:2019hho, Jho:2020jsa, Liu:2020qgx, Cline:2020mdt}.}

In the SM the dominant operator at the quark level that induces $K\to\pi2\nu$ has the form ${\cal C}\, \overline{s}\gamma^\mu(1-\gamma_5)d\, \overline{\nu}\gamma_\mu\nu+\rm H.c.$ with a complex coefficient $\cal C$ and a left-handed neutrino  field $\nu$.
The resulting interaction gives rise to an isospin change $\Delta I = 1/2$ and translates into the amplitude ratio $A(K_L\to\pi^0\nu\bar\nu)_{\rm SM}/A(K^+\to\pi^+\nu\bar\nu)_{\rm SM} = (\hat{\textsc a}_0/\hat{\textsc a}_+)({\rm Im}\,{\cal C})/{\cal C}$, with the factor $\hat{\textsc a}_0/\hat{\textsc a}_+\sim1$ manifesting approximate isospin symmetry, and the branching-fraction ratio $r_{\cal B}^{\rm SM}\simeq0.36$.
If NP is present and generates mainly or purely $\Delta I = 1/2$ effects on these modes, $r_{\cal B}$ may be modified and $K_L\to\pi^02\nu$, which is mostly $CP$-violating in the SM, may receive $CP$-conserving contributions.
Thus, such NP could raise $r_{\cal B}$ above $r_{\cal B}^{\rm SM}$ up to $r_{\cal B}^{\rm max}\simeq4.3$, which is largely due to the ratio $\tau_{K_L}/\tau_{K^+}\simeq4.1$ of the measured $K_L$ and $K^+$ lifetimes~\cite{Tanabashi:2018oca}. 
This can occur in many NP models which contribute to $K\to\pi2\nu$ via dimension-six operators involving the quark bilinears $\bar s\gamma^\mu(1\pm\gamma_5)d$ and SM fields~\cite{Grossman:1997sk,Buras:2004uu}.

Here we adopt the framework of effective field theory below the electroweak scale, where all the effective operators respect the SM residual gauge symmetry U(1)$_{\rm em}\times\rm SU(3)_{\rm color}$ and contain only the light fields of the minimal SM.
The heavier fields $c,b,W,Z,h$, and $t$ having been integrated out, we will dwell on sub-GeV interactions to examine how the GN bound can be violated.
Our focus is on operators which directly contribute to $K\to\pi2\nu$ and neglect long-distance contributions arising from Feynman diagrams mediated by light particles (charged leptons and/or mesons), as their effects are severely suppressed within and beyond the SM~\cite{Hagelin:1989wt,Rein:1989tr,Buchalla:1998ux,Lu:1994ww,Li:2019fhz}.
Given that the neutrinos in the final states are not experimentally identified and emerge as invisible particles, they can be a neutrino-antineutrino ($\nu\bar\nu$) pair if lepton number is conserved in the process or a pair of neutrinos ($\nu\nu$) or antineutrinos ($\bar\nu\bar\nu$) if lepton number is violated by two units. We will consider all these possibilities in our model-independent analysis.
Working with the mesonic realization of quark-level $\Delta I=1/2$ operators and concentrating on the pertinent kaon and pion interaction terms with the neutrinos, we demonstrate that the GN bound is always respected independent of the $CP$ property of the $\Delta I=1/2$ operators and whether the emitted neutrino pair has a zero or net lepton number. We also show in a general context that the presence of $\Delta I=3/2$ operators can lead to a violation of the GN bound, as already pointed out in Ref.\,\cite{He:2018uey} in a particular instance.

Isospin symmetry played an important role in the derivation of the GN bound.
If the relevant interactions originate from $\Delta I=1/2$ and $\Delta I=3/2$ quark-neutrino operators, they bring about, respectively, the components $A^{\Delta I=1/2}(K\to\pi 2\nu)$ and $A^{\Delta I=3/2}(K\to\pi 2\nu)$ in the decay amplitudes.
For the neutral and charged modes, they satisfy the relations
\begin{eqnarray}
{A^{\Delta I=1/2}(K^0 \to \pi^02\nu)\over A^{\Delta I=1/2}(K^+\to\pi^+2\nu)} \,=\, {-1\over \sqrt{2}} \,, ~~~~ ~~~ {A^{\Delta I=3/2}(K^0\to\pi^02\nu)\over A^{\Delta I=3/2}(K^+\to\pi^+2\nu)} \,=\, \sqrt{2} \;.
\end{eqnarray}
The GN bound is based on the assumption that the $\Delta I = 3/2$ interaction is absent.\footnote{We ignore the small effects of isospin breaking caused by the $u$- and $d$-quark mass difference and electroweak radiative corrections.}
Therefore, in the presence of the latter it would be possible to violate the bound.

Even if only the $\Delta I=1/2$ operators are present, but they involve more than two quarks besides the neutrinos and yield contributions with different $CP$ properties to $K_L\to\pi^02\nu$, they may seemingly undergo interference which makes its decay rate disrespect the GN bound. However, we find that this interference does not happen if $K^0$-$\bar K^0$ mixing, which is of order $2\times10^{-3}$ in size~\cite{Tanabashi:2018oca}, is neglected. We conclude that to violate the GN bound at an observable level requires the $\Delta I=3/2$ interactions to exist.  

\section{Operators contributing to \boldmath$K\to\pi\nu\bar\nu,\pi\nu\nu,\pi\bar\nu\bar\nu$}

An operator which has a two-quark part comprising just the $d$ and $s$ quarks can give rise to only $\Delta I= 1/2$ transitions directly. If the operator involves additional quarks, it may have both $\Delta I= 1/2$ and $\Delta I= 3/2$ components.
This applies to the $|\Delta S|=1$ local quark-level operators directly responsible for the lepton number conserving (LNC) decays $K\to\pi\nu_\alpha\bar{\nu}_\beta$ or the lepton number violating (LNV) ones $K\to\pi\nu_\alpha\nu_\beta,\pi\bar{\nu}_\alpha\bar{\nu}_\beta$, where $\alpha$ and $\beta$ refer to the neutrinos' flavors.
To investigate how such operators affect the $\Delta I=1/2,3/2$ amplitudes for these processes, we can work with the corresponding local effective operators involving the kaon, pion and neutrinos as dynamical degrees of freedom in the context of chiral perturbation theory~\cite{Gasser:1983yg,Gasser:1984gg,Kambor:1989tz}.

In the limit that isospin symmetry is preserved, the hadronic counterpart of a quark-level $\Delta I =1/2$ operator $\calO_{\Delta I=1/2}$ which induces  $K\to\pi\nu_\alpha\bar{\nu}_\beta$ or $K\to\pi\nu_\alpha\nu_\beta,\pi\bar{\nu}_\alpha\bar{\nu}_\beta$ can be expressed in the generic form\footnote{Implicitly, this belongs as usual to the Lagrangian \,$\hat c\,\calO_{\Delta I=1/2}+\rm H.c.$\, with $\hat c$ denoting a coupling constant of the appropriate mass dimension.\medskip}
\begin{eqnarray}\label{opebasis}
\bigg(\pi^- \partial_{\mu_1}\partial_{\mu_2}...\partial_{\mu_r}K^+ - \frac{1}{\sqrt2}\, \pi^0 \partial_{\mu_1}\partial_{\mu_2}...\partial_{\mu_r}K^0  \bigg) N_{\alpha\beta}^{\mu_1\mu_2\cdots\mu_r}
\end{eqnarray}
where $r\ge1$ and $N_{\alpha\beta}^{\mu_1\cdots\mu_r}$ stands for a neutrino current which can include derivatives.
Since the mesonic factor here is totally symmetric in its indices $\mu_1\cdots\mu_r$,
so is $N_{\alpha\beta}^{\mu_1\cdots\mu_r}$, any portions of the latter antisymmetric in any two of these indices having dropped out.
In Eq.\,(\ref{opebasis}) we have arranged the derivatives so that they all act on the kaon and neutrino fields but not on the pion ones.
This can always be achieved by (repeatedly) performing integration by parts and employing the particles' equations of motion, as we outline later on.
If the interaction changes isospin by $\Delta I = 3/2$ instead, the factor $-1/\sqrt{2}$ in front of \,$\pi^0\partial_{\mu_1}...\partial_{\mu_r}K^0$\, above is to be replaced by \,$+\sqrt{2}$.
In general, the $\Delta I=1/2,3/2$ contributions can be present simultaneously.

These local meson-neutrino operators can be classified according to their mass dimension and the property of the neutrino current. In this paper we restrict the operators to those with purely left-handed neutrinos and leave the right-handed neutrino case to a future publication.\footnote{Within specific models this case has been considered in \cite{He:2018uey,Hayreter:2019dzc}.}
Given that the kaon and pion fields each have mass dimension one, while the neutrino pair has mass dimension three, it is straightforward to realize that Lorentz invariance dictates the lowest dimension of the possible operators in the LNC and LNV cases to be~6 and~5, respectively.
For higher-dimensional operators, the dimension counting needs to take into account the contribution of the derivatives in them. It follows that the dimensions of LNC (LNV) operators are even (odd).
Moreover, with $n_f$ neutrino flavors, there are in total $(n-2)n_f^2$ independent LNC operators of the form displayed in Eq.~\eqref{opebasis} at dimension $2n$.
On the other hand, the total number of independent LNV operators is $(n-2)n_f^2+{1\over 2}[1-(-1)^n]n_f$ at dimension $2n-1$.
We now describe the expressions for these operators.

\subsection{LNC case\label{lnccase}}

Effective meson-neutrino operators which contribute to $K\to\pi 2\nu$ and conserve lepton number have mass dimension $2n\ge 6$ and can each be written as a linear combination of
\begin{eqnarray}\label{opebasislnc}
Q_{2n, m}^{\pm,\alpha\beta} &=& i \bigg( \pi^- \partial_{\mu_1}...\partial_{\mu_m}\partial_\rho K^+ - \frac{1}{\sqrt2}\, \pi^0\partial_{\mu_1}...\partial_{\mu_m}\partial_\rho K^0 \bigg) \partial^{2(n-3-m)}J_{\pm,\alpha\beta}^{\mu_1...\mu_m\rho},
\end{eqnarray}
where $m\le n-3$, with $m=0$ assigned to the case where the partial derivatives $\partial_{\mu_1}...\partial_{\mu_m}$ are absent, and\footnote{If a neutrino current has three gamma matrices, it can be rewritten in terms of currents with one gamma matrix with the aid of the identity $\gamma^\rho\gamma^\tau\gamma^\omega = g^{\rho \tau} \gamma^\omega + g^{\tau \omega} \gamma^\rho - g^{\omega \rho} \gamma^\tau - i\varepsilon^{\rho\tau\omega\mu} \gamma_\mu \gamma_5$ for $\varepsilon_{0123}=+1$. The remaining parts with the $\varepsilon$ tensor can be manipulated with the same identity and simplified into terms with one gamma matrix using (repeated) integration by parts and the neutrino equations of motion.}
\begin{eqnarray}\label{lnc:nucurrent}
J_{\pm,\alpha\beta}^{\mu_1...\mu_m\rho} &=& {1\over(1+\delta_{\alpha\beta})(1+\delta_{0m})}\Big[ \overline{\nu_\alpha}\gamma^\rho\partial^{\mu_1}...\partial^{\mu_m}\nu_\beta
+(-1)^m\, \overline{\mbox{$\partial^{\mu_1}...\partial^{\mu_m}$}\nu_\alpha}\gamma^\rho\nu_\beta \Big]\pm{\Hc}\,.
\end{eqnarray}
In this formula the factor with $\delta_{\alpha\beta}$ and $\delta_{0m}$ in the denominator has been added to ensure that there is no double-counting of these currents when the neutrinos have the same flavor, $\alpha=\beta$, and when $\partial_{\mu_1}...\partial_{\mu_m}$ are absent.
More specifically, in the latter case we set $m=0$, for which $J_{\pm,\alpha\beta}^{\mu_1...\mu_m\rho}$ become $J_{\pm,\alpha\beta}^{\rho}=\overline{\nu_\alpha}\gamma^\rho\nu_\beta/(1+\delta_{\alpha\beta})\pm{\rm H.c.}$
From Eq.\,(\ref{lnc:nucurrent}), we further see that
\begin{align}
\big(J_{\pm,\alpha\beta}^{\mu_1\cdots\mu_m\rho}\big)^\dagger &\,=\, \pm J_{\pm,\alpha\beta}^{\mu_1\cdots\mu_m\rho} , &
J_{\pm,\beta\alpha}^{\mu_1\cdots\mu_m\rho} &\,=\, \pm (-1)^m J_{\pm,\alpha\beta}^{\mu_1\cdots\mu_m\rho} ,
\end{align}
and consequently from the latter
$Q_{2n, m}^{\pm,\beta\alpha} \,=\, \pm (-1)^m Q_{2n, m}^{\pm,\alpha\beta}$.
It follows that $J_{+(-),\alpha\alpha}^{\mu_1\cdots\mu_m\rho}=0$ and hence $Q_{2n, m}^{+(-),\alpha\alpha}=0$ when $m$ is odd (even).
Upon comparing Eqs.\,\,(\ref{opebasis}) and (\ref{opebasislnc}), one can immediately identify
$\partial^{2(n-3-m)}J_{\pm,\alpha\beta}^{\mu_1\cdots\mu_m\rho}$ in $Q_{2n, m}^{\pm,\alpha\beta}$
as the neutrino currents $N^{\mu_1\cdots\mu_m\rho}$.

For example, the dimension-6 operator $\overline{s_L}\gamma_\mu d_L\overline{\nu_\alpha}\gamma^\mu\nu_\beta$, which occurs in many models (such as the SM if $\alpha=\beta$), corresponds to a combination of meson-neutrino operators with $n=3$ and $m=0$ in the notation convention of Eq.\,(\ref{opebasislnc}), namely
\begin{align}\label{lnc:example1}
Q_{6, 0}^{+,\alpha\beta}+Q_{6, 0}^{-,\alpha\beta} &\,=\, 2i \bigg( \pi^-\partial_\rho K^+ - \frac{1}{\sqrt2}\, \pi^0\partial_\rho K^0 \bigg)\overline{\nu_\alpha}\gamma^\rho\nu_\beta \,,
\end{align}
where the right-hand side results from employing chiral perturbation theory at leading order and subsequently applying integration by parts and the particles' equations of motion~\cite{Li:2019fhz}.
As another example, the dimension-10 operator $\big(\overline{s_R}\gamma_\mu u_R\, \overline{u_R}\gamma_\rho d_R + \overline{s_R}\gamma_\mu d_R\, \overline{d_R}\gamma_\rho d_R\big) \big(\overline{\nu_\alpha} i\mbox{$\stackrel{\leftrightarrow}{\partial}$}{}^\mu\gamma^\rho \nu_\alpha\big)$ corresponds to  $n=4$ and $m=1$ and hence
\begin{align}
Q_{8, 1}^{-,\alpha\alpha} &\,=\,  \bigg( \pi^-\partial_\mu\partial_\rho K^+ - \frac{1}{\sqrt2}\, \pi^0\partial_\mu\partial_\rho K^0 \bigg) \big(\overline{\nu_\alpha} i\mbox{$\stackrel{\leftrightarrow}{\partial}$}{}^\mu\gamma^\rho \nu_\alpha \big) \,, &
\end{align}
whereas $Q_{8, 1}^{+,\alpha\alpha}=0$.

\subsection{LNV case\label{lnvcase}}

Meson-neutrino operators contributing to $K\to\pi 2\nu$ that do not conserve lepton number have mass dimension $2n-1$, with $n\ge3$, and can each be written as a linear combination of
\begin{eqnarray}\label{opebasislnv}
Q_{2n-1, m}^{\pm,\alpha\beta} &=& \bigg(\pi^-  \partial_{\mu_1}...\partial_{\mu_m}K^+ - \frac{1}{\sqrt2}\, \pi^0\partial_{\mu_1}...\partial_{\mu_m}K^0 \bigg) \partial^{2(n-3-m)}j^{\mu_1...\mu_m}_{\pm,\alpha\beta} ,
\end{eqnarray}
where
\begin{eqnarray}\label{lnv:nucurrent}
j^{\mu_1...\mu_m}_{\pm,\alpha\beta} &=& {1\over1+\delta_{0m}} \Big[ \overline{\nu_\alpha}\, \partial^{\mu_1}...\partial^{\mu_m} \nu_\beta^{\textsc c} + (-1)^m\, \overline{\mbox{$\partial^{\mu_1}...\partial^{\mu_m}$}\nu_\alpha}\, \nu_\beta^{\textsc c} \Big]\pm{\Hc} ,
\end{eqnarray}
the superscript {\textsc c} indicating charge conjugation.\footnote{Generally speaking, neutrino tensor currents with $\sigma^{\mu\rho}=i[\gamma^\mu,\gamma^\rho]/2$ could also appear. However, with the aid of the identity $i\sigma^{\mu\rho}=g^{\mu\rho}-\gamma^\mu\gamma^\rho=\gamma^\rho\gamma^\mu-g^{\mu\rho}$ and the particles' equations of motion, they can be reduced to currents without gamma matrices as in Eq.\,(\ref{lnv:nucurrent}).}
In the absence of the partial derivatives $\partial^{\mu_1}...\partial^{\mu_m}$, we set $m=0$ and so the currents in Eq.\,(\ref{lnv:nucurrent}) become $j_{\pm,\alpha\beta}=\overline{\nu_\alpha}\nu_\beta^{\textsc c}\pm\overline{\nu_\alpha^{\textsc c}}\nu_\beta$.
Clearly, $j^{\mu_1...\mu_m}_{\pm,\alpha\beta}$ change lepton number by two units and
\begin{align}
\big(j^{\mu_1...\mu_m}_{\pm,\alpha\beta}\big)^\dagger &\,=\,
\pm j^{\mu_1...\mu_m}_{\pm,\alpha\beta} \,, &
j^{\mu_1...\mu_m}_{\pm,\beta\alpha} &\,=\, (-1)^m\, j^{\mu_1...\mu_m}_{\pm,\alpha\beta} \,, ~~~
\end{align}
and consequently from the latter $Q_{2n-1, m}^{\pm,\beta\alpha}=(-1)^m\, Q_{2n-1, m}^{\pm,\alpha\beta}$, implying that $Q_{2n-1, m}^{\pm,\alpha\alpha}=0$ when $m$ is odd.
For instance, the dimension-6 and -7 operators $\overline{s_L}d_R\, \overline{\nu_\alpha} \nu_\beta^{\textsc c}$ and $\overline{s_L}\gamma_\mu d_L \big(\overline{\nu_\alpha} i\mbox{$\stackrel{\leftrightarrow}{\partial}$}{}^\mu\nu_\beta^{\textsc c}\big)$ correspond, respectively, to
\begin{align}
Q_{5,0}^{+,\alpha\beta} + Q_{5,0}^{-,\alpha\beta} &\,=\, 2 \bigg( \pi^- K^+ - \frac{1}{\sqrt2}\, \pi^0K^0 \bigg) \overline{\nu_\alpha}\nu_\beta^{\textsc c} \,,
\\
Q_{7,1}^{+,\alpha\beta} + Q_{7,1}^{-,\alpha\beta} &\,=\, 2 \bigg( \pi^-\partial_\mu K^+ - \frac{1}{\sqrt2}\, \pi^0\partial_\mu K^0 \bigg) \big( \overline{\nu_\alpha} \mbox{$\stackrel{\leftrightarrow}{\partial}$}{}^\mu\nu_\beta^{\textsc c} \big) \,.
\end{align}

\subsection{Completeness and independence of operators\label{complete}}

The above LNC and LNV meson-neutrino operators are independent and form a complete operator basis.
Other operators can be expressed as linear combinations of those in this basis by means of the following relations.

Firstly, an operator with a neutrino-current factor $\partial^\mu N^{\mu_1...\mu_m}$ can be transformed into an operator having fewer derivatives and another involving $\partial^2 N^{\mu_1...\mu_m}$ after the application of integration by parts and the equations of motion $\partial^2K=m_K^2K$ and $\partial^2\pi=m_\pi^2\pi$.
Thus
\begin{eqnarray}\label{reduce:relation1}
\pi \big( \partial_\mu\partial_{\mu_1}...\partial_{\mu_m}K \big) \partial^\mu N^{\mu_1...\mu_m} &=& -{1\over 2} \big( m_K^2\, \pi\, \partial_{\mu_1}...\partial_{\mu_m} K
-m_\pi^2\, \pi\, \partial_{\mu_1}...\partial_{\mu_m}K \big) N^{\mu_1...\mu_m}
\nonumber \\ && \!
-\; {1\over 2}\pi \big( \partial_{\mu_1}...\partial_{\mu_m}K \big) \partial^2N^{\mu_1...\mu_m} ,
\end{eqnarray}
where the first term on the right-hand side is of the form in Eq.~\eqref{opebasis} and the second term has the form in  Eq.~\eqref{opebasislnc} or \eqref{opebasislnv}.

Secondly, in the neutrino part of each operator the derivatives can be arranged so that all of them act on only one of the neutrino fields.
For example, in the LNC and LNV cases we could have, respectively,
\begin{eqnarray}\label{reduce:relation2}
\overline{\partial^\mu\nu_\alpha}\gamma^\rho \partial^{\mu_{1}}...\partial^{\mu_m} \nu_\beta
&=& \partial^\mu N_{\alpha\beta}^{\mu_1...\mu_m\rho} - \overline{\nu_\alpha}\gamma^\rho  \partial^\mu\partial^{\mu_{1}}...\partial^{\mu_m} \nu_\beta  \,,
\nonumber \\
\overline{\partial^\mu\nu_\alpha}\partial^{\mu_{1}}...\partial^{\mu_m} \nu_\beta^{\textsc c}
&=& \partial^\mu N_{\alpha\beta}^{\mu_1...\mu_m} - \overline{\nu_\alpha}\partial^\mu\partial^{\mu_{1}}...\partial^{\mu_m} \nu_\beta^{\textsc c} \,,
\end{eqnarray}
where the expressions for $N_{\alpha\beta}^{\mu_1...\mu_m\rho}$ and $N_{\alpha\beta}^{\mu_1...\mu_m}$ are not explicitly displayed but can be easily written down. Upon contracting the last two equations with the meson parts, we would see that in the resulting operators the terms containing the $N$s are just of the form in Eq.~\eqref{reduce:relation1} and the rest of the terms are as those with the currents in Eq.~\eqref{lnc:nucurrent} or~\eqref{lnv:nucurrent}.

Thirdly, these neutrino currents can be further arranged to be symmetric or antisymmetric under the interchange of $\alpha$ and $\beta$ with the aid of
\begin{eqnarray}\label{reduce:relation3}
\overline{\nu_\alpha}\gamma^\rho\partial^{\mu_{1}}...\partial^{\mu_m} \nu_\beta
&=&(-1)^m\, \overline{\partial^{\mu_{1}}...\partial^{\mu_m}\nu_\alpha}\gamma^\rho \nu_\beta
+ \boxed{\partial^{\mu_1}N_{\alpha\beta}^{\mu_2...\mu_m}+...} ,
\nonumber \\
\overline{\nu_\alpha}\partial^{\mu_{1}}...\partial^{\mu_m} \nu_\beta^{\textsc c}
&=&(-1)^m \overline{\nu_\beta}\partial^{\mu_{1}}...\partial^{\mu_m}\nu_\alpha^{\textsc c}+\boxed{\partial^{\mu_1}N_{\alpha\beta}^{\mu_2...\mu_m}+...} ,
\end{eqnarray}
where the boxed parts contain a series of terms with derivatives acting on various $N$s.
Lastly, each neutrino current can always be decomposed as a sum of Hermitian and anti-Hermitian components, as was already done in Eqs.\,\,\eqref{lnc:nucurrent} and~\eqref{lnv:nucurrent}.

\section{\boldmath LNC amplitudes from $\Delta I= 1/2$ interaction only\label{LNCA}}

We can express the interaction Lagrangian containing the LNC operators $Q_{2n, m}^{\pm,\alpha\beta}$ in Eq.\,\eqref{opebasislnc} as
\begin{eqnarray} \label{Llnc} {\cal L}_{\pi K2\nu}^{\textsc{lnc}} &=& \sum_{m,n} \Big( {\cal C}_{2n,m}^{+,\alpha\beta}Q_{2n, m}^{+,\alpha\beta} \,+\, {\cal C}_{2n, m}^{-,\alpha\beta}Q_{2n, m}^{-,\alpha\beta} \Big) \,+\, {\rm H.c.} ,
\end{eqnarray}
where we have taken into account the contributions of all possible LNC operators with $n\ge 3$ and $0\le m\le n-3$ and ${\cal C}_{2n, m}^{\pm,\alpha\beta}$ are generally complex coefficients which have mass dimension $4-2n$. Since $Q_{2n, m}^{\pm,\beta\alpha}=\pm(-1)^mQ_{2n, m}^{\pm,\alpha\beta}$, as pointed out in subsection\,\,\ref{lnccase}, it is unnecessary to include the terms ${\cal C}_{2n,m}^{+,\beta\alpha}Q_{2n, m}^{+,\beta\alpha}+{\cal C}_{2n, m}^{-,\beta\alpha}Q_{2n, m}^{-,\beta\alpha}$ in ${\cal L}_{\pi K2\nu}^{\textsc{lnc}}$ because they would only lead to the redefining of ${\cal C}_{2n, m}^{\pm,\alpha\beta}$.

For the decays $K^+(k)\to\pi^+\nu_\alpha(p)\bar{\nu}_\beta(\overline p)$ and $K_{L}(k)\to\pi^0\nu_\alpha(p)\bar{\nu}_\beta(\overline p)$, it is then straightforward to derive from Eq.\,(\ref{Llnc}), in conjunction with the approximate relations $\sqrt2\,K^0=(1-\epsilon)(K_L+K_S)$ and $\sqrt2\,\bar K^0=(1+\epsilon)(K_L-K_S)$, the amplitudes
\begin{eqnarray} \label{lncMK2pnn}
\mathcal{A}_{K^+\to\pi^+\nu_\alpha\bar{\nu}_\beta} &=& \sum_{m,n}
\frac{{\cal C}_{2n,m}^{+,\alpha\beta} + {\cal C}_{2n,m}^{-,\alpha\beta}}{1+\delta_{0m}} (-\hat s)^{n-3-m} \big[ (k\cdot\overline p)^m+(-k\cdot p)^m \big] \overline{u_\alpha}\slashed{k}P_Lv_{\beta}
\nonumber \\ &=& \big( A_+^{\alpha\beta} + A_-^{\alpha\beta} + B_+^{\alpha\beta} + B_-^{\alpha\beta}\big) \overline{u_\alpha}\slashed{k}P_Lv_\beta \,, \vphantom{|_{|_|^|}}
\\ \label{lnc'MK2pnn}
\mathcal{A}_{K_L\to\pi^0\nu_\alpha\bar{\nu}_\beta} &=& \sum_{m,n}
\frac{-i{\rm Im}\,{\cal C}_{2n,m}^{+,\alpha\beta} - {\rm Re}\,{\cal C}_{2n,m}^{-,\alpha\beta}}{1+\delta_{0m}} (-\hat s)^{n-3-m} \big[ (k\cdot\overline p)^m+(-k\cdot p)^m \big] \overline{u_\alpha}\slashed{k}P_Lv_\beta
\nonumber \\ &=& -\big( i{\rm Im}\, A_+^{\alpha\beta} + {\rm Re}\, A_-^{\alpha\beta} + i{\rm Im}\, B_+^{\alpha\beta} + {\rm Re}\,  B_-^{\alpha\beta} \big) \overline{u_\alpha}\slashed{k}P_Lv_\beta \,,
\vphantom{|_{|_|^|}} 
\end{eqnarray}
where $u_\alpha$ and $v_{\beta}$ are the Dirac spinors of $\nu_\alpha$ and $\bar\nu_\beta$, respectively, $\hat s=(p+\overline p)^2$ is the 2$\nu$ invariant mass squared, $P_L=(1-\gamma_5)/2$,
\begin{eqnarray}\label{lnc:amp}
A^{\alpha\beta}_\pm &=& \sum_{n=3}\, \sum_{m \in \rm even} {{\cal C}_{2n,m}^{\pm,\alpha\beta} \over 1+\delta_{0m}} (-\hat s)^{n-3-m} \big[ (k\cdot\overline p)^m+(k\cdot p)^m \big] \,,
\nonumber \\
B^{\alpha\beta}_\pm &=& \sum_{n=4}\, \sum_{m \in \rm odd}{\cal C}_{2n,m}^{\pm,\alpha\beta} (-\hat s)^{n-3-m}\big[ (k\cdot\overline p)^m-(k\cdot p)^m \big] \,,
\end{eqnarray}
and we have dropped terms proportional to the kaon-mixing parameter $\epsilon$ which is around $2\times10^{-3}$ in size~\cite{Tanabashi:2018oca}.
If $\alpha=\beta$ in these amplitudes, we need to impose ${\cal C}_{2n,m}^{+(-),\alpha\alpha}=0$ when $m$ is odd (even) because, as mentioned in the preceding section, $Q_{2n, m}^{+(-),\alpha\alpha}=0$ when $m$ is odd (even).
Accordingly, $A^{\alpha\alpha}_-=0$ and $B^{\alpha\alpha}_+=0$.
The SM contributes to the $\alpha=\beta$ amplitudes in Eqs.\,\,(\ref{lncMK2pnn})-(\ref{lnc'MK2pnn}), in the terms with $2n=6$ and $m=0$.
For $\alpha\neq\beta$, from ${\cal L}_{\pi K2\nu}^{\textsc{lnc}}$ there are the extra modes $K^+(k)\to\pi^+\nu_\beta(p)\bar{\nu}_\alpha(\overline p)$ and $K_{L}(k)\to\pi^0\nu_\beta(p)\bar{\nu}_\alpha(\overline p)$, for which the amplitudes are
\begin{eqnarray} \label{lncMK2pnn'}
\mathcal{A}_{K^+\to\pi^+\nu_\beta\bar\nu_\alpha} &=& \sum_{m,n}
\frac{{\cal C}_{2n,m}^{+,\alpha\beta} - {\cal C}_{2n,m}^{-,\alpha\beta}}{1+\delta_{0m}} (-\hat s)^{n-3-m} \big[ (k\cdot p)^m+(-k\cdot\overline p)^m \big] \overline{u_\beta}\slashed k P_L v_\alpha
\nonumber \\ &=& \big( A_+^{\alpha\beta} - A_-^{\alpha\beta} + B_+^{\alpha\beta} - B_-^{\alpha\beta} \big) \overline{u_\alpha}\slashed{k}P_Lv_\beta \,, \vphantom{|_{|_|^|}}
\\ \label{lnc'MK2pnn'}
\mathcal{A}_{K_L\to\pi^0\nu_\beta\bar\nu_\alpha} &=& \sum_{m,n}
\frac{-i{\rm Im}\,{\cal C}_{2n,m}^{+,\alpha\beta} + {\rm Re}\,{\cal C}_{2n,m}^{-,\alpha\beta}}{1+\delta_{0m}} (-\hat s)^{n-3-m} \big[ (k\cdot p)^m+(-k\cdot\overline p)^m \big] \overline{u_\beta} \slashed k P_L v_\alpha
\nonumber \\ &=& -\big( i{\rm Im}\, A_+^{\alpha\beta} - {\rm Re}\, A_-^{\alpha\beta} + i{\rm Im}\, B_+^{\alpha\beta} - {\rm Re}\,  B_-^{\alpha\beta} \big) \overline{u_\alpha}\slashed{k}P_Lv_\beta \,.
\end{eqnarray}

To evaluate the rate from the spin-summed absolute square of each amplitude $\mathcal{A}_{K\to\pi\nu\bar\nu'}$, we first find $\sum_{\rm spins}|\overline{u_\nu}\slashed{k}P_Lv_{\nu'}|^2=4(k\cdot p)(k\cdot\overline p)-m_K^2\hat s$, neglecting neutrino masses, and use it to define the phase-space factor
\begin{eqnarray}
d\widehat{\Pi}_3 &=& {d\Pi_3\over 2m_K}\, \sum_{\rm spins}|\overline{u_\nu}\slashed{k}P_Lv_{\nu'}|^2 \,=\, {d\Pi_3 \over 2m_K} \big[ 4(k\cdot p)(k\cdot\overline p)-m_K^2s \big] \,.
\end{eqnarray}
Then, from Eqs.\,\,(\ref{lncMK2pnn}), (\ref{lnc'MK2pnn}), (\ref{lncMK2pnn'}), and\,\,(\ref{lnc'MK2pnn'})  we arrive at
\begin{align} \label{lnc:decaywidth}
\Gamma\big(K^+\to\pi^+\nu_\alpha\bar\nu_\beta\big) &= \int d\widehat{\Pi}_3 \Big( \big|A_+^{\alpha\beta}+A_-^{\alpha\beta}\big|^2 + \big|B_+^{\alpha\beta}+ B_-^{\alpha\beta}\big|^2 \Big) \,,
\nonumber \\
\Gamma\big(K^+\to\pi^+\nu_\beta\bar\nu_\alpha\big) &= \int d\widehat{\Pi}_3 \Big( \big|A_+^{\alpha\beta} - A_-^{\alpha\beta}\big|^2 + \big|B_+^{\alpha\beta} - B_-^{\alpha\beta} \big|^2 \Big)  \,,
\nonumber \\ 
\Gamma\big(K_L\to\pi^0\nu_\alpha\bar\nu_\beta\big) &=\Gamma\big(K_L\to\pi^0\nu_\beta\bar\nu_\alpha\big)=\int d\widehat{\Pi}_3 \Big({\rm Im}^2A_+^{\alpha\beta} +{\rm Re}^2A_-^{\alpha\beta} +{\rm Im}^2B_+^{\alpha\beta} +{\rm Re}^2 B_-^{\alpha\beta} \Big),
\end{align}
in the isospin-symmetric limit, the $A_i^*B_j$ interference terms having vanished after phase-space integration due to their being antisymmetric under the exchange of $p$ and $\overline p$, as can be checked explicitly.
Evidently these rates fulfill the relations
\begin{eqnarray}\label{lnc:gnbound}
\Gamma\big(K_L\to\pi^0\nu_\alpha\bar\nu_\beta\big) + \Gamma\big(K_L\to\pi^0\nu_\beta\bar\nu_\alpha\big) &\le& \Gamma\big(K^+\to\pi^+\nu_\alpha\bar\nu_\beta\big) + \Gamma\big(K^+\to\pi^+\nu_\beta\bar\nu_\alpha\big) \,. ~~~~~
\end{eqnarray}
Since the neutrinos' flavors are not experimentally identified, both sides of this relation need to be summed over $\alpha$ and $\beta$.
With the $K_L$ and $K^+$ lifetimes included in Eq.\,(\ref{lnc:gnbound}), the resulting ratio of the $K_L$ and $K^+$ branching fractions reproduces the GN bound.
It is worth noting that we arrive at this conclusion without paying attention to the $CP$ properties of the responsible $\Delta I=1/2$ operators.\footnote{This differs from the conclusion drawn in section 5 of~\cite{He:2018uey} that the GN bound could be violated by $CP$ conserving effects. This difference is because of a missing imaginary unit $i$ in Eq.\,(C4) in Appendix C in~\cite{He:2018uey}, which implies that the resulting NP contribution to the $K_L$ amplitude should be purely imaginary as in the SM part and therefore Eq.\,(14) therein needs to be corrected accordingly.\medskip}
Furthermore, it is clearly independent of whether or not lepton flavor is conserved.\footnote{In \cite{Mandal:2019gff,Pich:2020gan} the authors claim that the GN bound can be violated if the emitted neutrinos have different flavors. However, applying the Cauchy-Schwarz inequality to the lepton-flavor-violating (LFV) part of eq.\,(5.22) in \cite{Mandal:2019gff} for BR$(K_L\to\pi^0\nu\bar\nu)$ results in $\sum_{m \neq n}\big|[N_{V_{X}}^{d}]^{21, m n}-[N_{V_{X}}^{d}]^{12, m n}|^2\leq 4\sum_{m \neq n} |[N_{V_{X}}^{d}]^{21, m n}|^2$. This implies that the LFV parts of eq.\,(5.21) for BR$(K^+\to\pi^+\nu\bar\nu)$ in \cite{Mandal:2019gff} and of eq.\,(5.22) therein obey the GN bound.}

\section{LNV amplitudes from \boldmath$\Delta I =1/2$ interaction only\label{LNVA}}

We can write the effective Lagrangian containing the LNV operators $Q_{2n-1,m}^{\pm,\alpha\beta}$ in Eq.\,\eqref{opebasislnv} as
\begin{eqnarray} \label{Llnv}
{\cal L}_{\pi K2\nu}^{\textsc{lnv}} &=& \sum_{m,n} \Big( {\cal C}_{2n-1,m}^{+,\alpha\beta}Q_{2n-1, m}^{+,\alpha\beta} \,+\, {\cal C}_{2n-1, m}^{-,\alpha\beta}Q_{2n-1, m}^{-,\alpha\beta} \Big) \,+\, {\rm H.c.}
\end{eqnarray}
where we have included all possible LNV operators with various $m$ and $n$ values and ${\cal C}_{2n-1,m}^{\pm,\alpha\beta}$ are generally complex coefficients of mass dimension $5-2n$ and encode the underlying NP.
Since $Q_{2n-1, m}^{\pm,\beta\alpha}=(-1)^m\, Q_{2n-1, m}^{\pm,\alpha\beta}$ according to subsection\,\,\ref{lnvcase}, adding to Eq.\,(\ref{Llnv}) the extra terms
${\cal C}_{2n-1,m}^{+,\beta\alpha}Q_{2n-1, m}^{+,\beta\alpha} + {\cal C}_{2n-1, m}^{-,\beta\alpha}Q_{2n-1, m}^{-,\beta\alpha}$
would only amount to redefining ${\cal C}_{2n-1, m}^{\pm,\alpha\beta}$.

From ${\cal L}_{\pi K2\nu}^{\textsc{lnv}}$, we derive the amplitudes for the LNV decays $K(k)\rightarrow\pi\nu_\alpha(p)\nu_\beta(\overline p)$ to be
\begin{eqnarray} \label{lnv2nu:totalamp}\nonumber
{\cal A}_{K^+\to\pi^+\nu_\alpha\nu_\beta} &=& \sum_{m,n} {{\cal C}_{2n-1,m}^{+,\alpha\beta}+{\cal C}_{2n-1,m}^{-,\alpha\beta}\over 1+\delta_{0m}}(-\hat s)^{n-3-m}
\big[ (k\cdot\overline p)^m+(-k\cdot p)^m \big] \overline{u_\alpha}P_R u_\beta^{\textsc c}
\\
&=& \big( C_+^{\alpha\beta}+C_-^{\alpha\beta} \big) \overline{u_\alpha}P_R u_\beta^{\textsc c} \,,
\\ \label{lnv2nu':totalamp} \nonumber
{\cal A}_{K_L\to\pi^0\nu_\alpha\nu_\beta} &=& \sum_{m,n}
{ -{\rm Re}\, {\cal C}_{2n-1,m}^{+,\alpha\beta}-i{\rm Im}\, {\cal C}_{2n-1,m}^{-,\alpha\beta}\over 1+\delta_{0m}}(-\hat s)^{n-3-m} \big[ (k\cdot\overline p)^m+(-k\cdot p)^m \big] \overline{u_\alpha}P_Ru_\beta^{\textsc c}
\\ &=& -\big( {\rm Re}\, C_+^{\alpha\beta}+i{\rm Im}\, C_-^{\alpha\beta} \big) \overline{u_\alpha}P_R u_\beta^{\textsc c} \,,
\end{eqnarray}
where $u_\alpha$ and $u_\beta$ are the Dirac spinors of $\nu_\alpha$ and $\nu_\beta$, respectively, $P_R=(1+\gamma_5)/2$, and
\begin{eqnarray} \label{lnv:amp}
C^{\alpha\beta}_\pm &=& \sum_{n=3}\, \sum_{m=0}^{n-3}{{\cal C}_{2n-1,m}^{\pm,\alpha\beta}\over 1+\delta_{0m}}(-\hat s)^{n-3-m} \big[ (k\cdot\overline p)^m+(-k\cdot p)^m \big] \,.
\end{eqnarray}
From ${\cal L}_{\pi K2\nu}^{\textsc{lnv}}$ we can also obtain the amplitudes for $K(k)\to\pi\bar{\nu}_\alpha(p)\bar{\nu}_\beta(\overline p)$, with two anti-neutrinos in the final states, namely
\begin{eqnarray} \label{lnv2antinu:totalamp'} \nonumber
{\cal A}_{K^+\to\pi^+\bar\nu_\alpha\bar\nu_\beta} &=& \sum_{m,n} {{\cal C}_{2n-1,m}^{+,\alpha\beta}-{\cal C}_{2n-1,m}^{-,\alpha\beta} \over 1+\delta_{0m}} (-\hat s)^{n-3-m} \big[ (k\cdot\overline p)^m+(-k\cdot p)^m \big] \overline{v_\alpha^{\textsc c}}P_Lv_\beta
\\
&=& \big( C_+^{\alpha\beta}-C_-^{\alpha\beta} \big) \overline{v_\alpha^{\textsc c}}P_Lv_\beta \,,
\\\label{lnv2antinu':totalamp'} \nonumber
{\cal A}_{K_L\to\pi^0\bar\nu_\alpha\bar\nu_\beta} &=& \sum_{m,n} {-{\rm Re}\, {\cal C}_{2n-1,m}^{+,\alpha\beta}+i{\rm Im}\, {\cal C}_{2n-1,m}^{-,\alpha\beta}\over 1+\delta_{0m}} (-\hat s)^{n-3-m} \big[ (k\cdot\overline p)^m+(-k\cdot p)^m \big] \overline{v_\alpha^{\textsc c}}P_Lv_\beta
\\ &=& -\big( {\rm Re}\, C_+^{\alpha\beta} - i{\rm Im}\, C_-^{\alpha\beta} \big) \overline{v_\alpha^{\textsc c}}P_Lv_\beta \,,
\end{eqnarray}
where $v_\alpha$ and $v_\beta$ are the Dirac spinors of $\bar\nu_\alpha$ and $\bar\nu_\beta$, respectively.

To evaluate the rates, we first obtain $\Sigma_{\rm spins}|\overline{u_\alpha}P_Ru_\beta^{\textsc c}|^2=\Sigma_{\rm spins}|\overline{v_\alpha^{\textsc c}}P_Lv_\beta|^2=\hat s$, neglecting neutrino masses, to get the phase-space factor
\begin{eqnarray}
d\widetilde{\Pi}_3 &=& {d\Pi_3\over 2m_K}\sum_{\rm spins} |\overline{u_\alpha}P_Ru_\beta^{\textsc c}|^2 \,=\, {d\Pi_3\over 2m_K}\sum_{\rm spins} |\overline{v_\alpha}P_Lv_\beta^{\textsc c}|^2 \,=\, {d\Pi_3\, \hat s\over 2m_K}\, .
\end{eqnarray}
Then, from Eqs.\,\,(\ref{lnv2nu:totalamp}), (\ref{lnv2nu':totalamp}), (\ref{lnv2antinu':totalamp'}), and\,\,(\ref{lnv2antinu':totalamp'})  we arrive at
\begin{align} \label{lnv:decaywidth} \nonumber
\Gamma\big(K^+\to\pi^+\nu_\alpha\nu_\beta\big) &={1 \over 1+\delta_{\alpha\beta}} \int d\widetilde{\Pi}_3 \big| C_+^{\alpha\beta} + C_-^{\alpha\beta} \big|^2 \,, 
\\\nonumber
\Gamma\big(K^+\to\pi^+\bar\nu_\alpha\bar\nu_\beta\big)&={1 \over 1+\delta_{\alpha\beta}}\int d\widetilde{\Pi}_3 \big| C_+^{\alpha\beta}-C_-^{\alpha\beta} \big|^2 \,,
\\
\Gamma\big(K_L\to\pi^0\nu_\alpha\nu_\beta\big) &=
\Gamma\big(K_L\to\pi^0\bar\nu_\alpha\bar\nu_\beta\big)=
{1 \over 1+\delta_{\alpha\beta}} \int d\widetilde{\Pi}_3 \Big({\rm Re}^2C_+^{\alpha\beta} + {\rm Im}^2C_-^{\alpha\beta} \Big) \,,
\end{align}
again in the isospin-symmetric limit, with the factor $1/(1+\delta_{\alpha\beta})$ in each equation accounting for the identical particles in the final state if $\alpha=\beta$. 
As the neutrino or antineutrino pair is not observed, these rates lead to
\begin{eqnarray} \label{lnv:gnbound}
\frac{\Gamma\big(K_L\to\pi^0\nu_\alpha\nu_\beta\big) + \Gamma\big(K_L\to\pi^0\bar\nu_\alpha\bar\nu_\beta\big)}{\Gamma\big(K^+\to\pi^+\nu_\alpha\nu_\beta\big) + \Gamma\big(K^+\to\pi^+\bar\nu_\alpha\bar\nu_\beta\big)} &\le& 1 \,.
\end{eqnarray}
The right-hand side of this equation is unchanged if the numerator and denominator on the left-hand side are both summed over the neutrino flavors. These relations are equivalent to the GN bound in the LNV case.

Given that the LNC and LNV contributions do not interfere with each other, combining Eqs.~\eqref{lnc:gnbound} and \eqref{lnv:gnbound}, we find the most general relation from purely $\Delta I=1/2$ interactions:
\begin{eqnarray} \label{gen}
{\Gamma(K_L\to\pi^0\nu_\alpha\bar\nu_\beta)+\Gamma(K_L\to\pi^0\nu_\beta\bar\nu_\alpha)+\Gamma(K_L\to \pi^0\nu_\alpha\nu_\beta)+\Gamma(K_L\to \pi^0\bar\nu_\alpha\bar\nu_\beta)
\over
\Gamma(K^+\to\pi^+\nu_\alpha\bar\nu_\beta)+\Gamma(K^+\to\pi^+\nu_\beta\bar\nu_\alpha)+\Gamma(K^+\to \pi^+\nu_\alpha\nu_\beta)+\Gamma(K^+\to \pi^+\bar\nu_\alpha\bar\nu_\beta) }\leq 1 \,.
\end{eqnarray}
Again, experimentally $\alpha$ and $\beta$ are summed over and this relation is still valid. 
Converting Eq.\,(\ref{gen}) to the ratio of branching fractions yields the GN bound $r_{\cal B}^{\Delta I=1/2}\leq r_{\cal B}^{\rm max}\simeq4.3$, irrespective of the two neutrinos conserving lepton number or not.

\section{General case with both \boldmath$\Delta I=1/2$ and $\Delta I=3/2$ interactions\label{comb}}

The analysis in the last two sections can be easily redone for the purely $\Delta I=3/2$ case, with $-1/\sqrt{2}$ in Eq.~\eqref{opebasis} replaced by $+\sqrt{2}$.  This results in
\begin{eqnarray}
\frac{\Gamma(K_L\to\pi^0\nu_\alpha\bar\nu_\beta)+\Gamma(K_L\to\pi^0\nu_\beta\bar\nu_\alpha)}{\Gamma(K^+\to\pi^+\nu_\alpha\bar\nu_\beta)+\Gamma(K^+\to\pi^+\nu_\beta\bar\nu_\alpha)} &\leq& 4 \,,
\nonumber \\
{\Gamma(K_L\to \pi^0\nu_\alpha\nu_\beta)+\Gamma(K_L\to \pi^0\bar\nu_\alpha\bar\nu_\beta) \over
\Gamma(K^+\to \pi^+\nu_\alpha\nu_\beta)+\Gamma(K^+\to \pi^+\bar\nu_\alpha\bar\nu_\beta)}
&\leq& 4 
\end{eqnarray}
in the LNC and LNV scenarios, respectively.
After incorporating the $K^+$ and $K_L$ lifetimes, the original GN bound $r_{\cal B}^{\Delta I=1/2}\leq 4.3$ is now modified to $r_{\cal B}^{\Delta I=3/2}\leq4.3\times4 \simeq 17$.

The situation described in the preceding paragraph is, of course, not realistic because the SM already generates $\Delta I = 1/2$ amplitudes.
The breaking of the GN bound is more likely the combined effect of $\Delta I = 1/2$ and $\Delta I =3/2$ operators.
We now extend the $\Delta I=1/2$ case discussed in the earlier sections to the more general case in which both $\Delta I=1/2$ and $\Delta I=3/2$ operators coexist. In the following we keep the notation ${\cal C}_i$ for the coefficients of the $\Delta I=1/2$ operators from Sects.\,\,\ref{LNCA} and\,\,\ref{LNVA}  and adopt $\tilde{\cal C}_i$ to denote the coefficients of the $\Delta I=3/2$ operators from the replacement $-1/\sqrt{2}\to \sqrt{2}$ in the LNC and LNV formulas in Eqs.\,\,\eqref{opebasislnc} and\,\,\eqref{opebasislnv}, respectively.
Correspondingly, $\big(\tilde A_i,\tilde B_i,\tilde C_i\big)$ are the $\Delta I=3/2$ counterparts of $(A_i, B_i, C_i)$  in Eqs.\,\,(\ref{lnc:decaywidth}) and\,\,(\ref{lnv:decaywidth}). Thus, the ratio of branching fractions of the LNC decays becomes
\begin{align}\label{lnc:ratioofbr}
& {\mathcal{B}(K_L\to\pi^0\nu_\alpha\bar{\nu}_\beta)+\mathcal{B}(K_L\to\pi^0\nu_\beta\bar{\nu}_\alpha) \over \mathcal{B}(K^+\to\pi^+\nu_\alpha\bar{\nu}_\beta)+\mathcal{B}(K^+\to\pi^+\nu_\beta\bar{\nu}_\alpha)}
= \frac{\tau_{K_L}}{\tau_{K^+}}~\times
\nonumber
\\ &
{\int d\widehat{\Pi}_3 \Big[ {\rm Im}^2\big( A^{\alpha\beta}_+-2\tilde{A}^{\alpha\beta}_+ \big) +{\rm Re}^2\big( A^{\alpha\beta}_--2\tilde{A}^{\alpha\beta}_- \big) +{\rm Im}^2\big( B^{\alpha\beta}_+-2\tilde{B}^{\alpha\beta}_+ \big) +{\rm Re}^2\big( B^{\alpha\beta}_--2\tilde{B}^{\alpha\beta}_- \big) \Big]\over
\int d\widehat{\Pi}_3 \Big( \big| A^{\alpha\beta}_++\tilde{A}^{\alpha\beta}_+\big|^2 +\big|A^{\alpha\beta}_-+\tilde{A}^{\alpha\beta}_-\big|^2 +\big|B^{\alpha\beta}_+ + \tilde{B}^{\alpha\beta}_+\big|^2 +\big|B^{\alpha\beta}_-+\tilde{B}^{\alpha\beta}_-\big|^2 \Big) }\,.
\end{align}
This can in general have any positive value if there is no requirement on the $\Delta I=1/2$ and $\Delta I=3/2$ components.
For the LNV transitions, we have
\begin{align}\label{lnv:ratioofbr}
& {\mathcal{B}(K_L\to\pi^0\nu_\alpha\nu_\beta)+\mathcal{B}(K_L\to\pi^0\bar{\nu}_\alpha\bar{\nu}_\beta) \over
\mathcal{B}(K^+\to\pi^+\nu_\alpha\nu_\beta)+\mathcal{B}(K^+\to\pi^+\bar{\nu}_\alpha\bar{\nu}_\beta)} = \frac{\tau_{K_L}}{\tau_{K^+}}
{\int d\widetilde{\Pi}_3 \Big[ {\rm Re}^2\big(C^{\alpha\beta}_+-2\tilde{C}^{\alpha\beta}_+\big)+{\rm Im}^2\big(C^{\alpha\beta}_--2\tilde{C}^{\alpha\beta}_-\big) \Big]
\over
\int d\widetilde{\Pi}_3 \Big( \big|C^{\alpha\beta}_++\tilde{C}^{\alpha\beta}_+\big|^2+\big|C^{\alpha\beta}_-+\tilde{C}^{\alpha\beta}_-\big|^2 \Big)} \,,
\end{align}
which leads to a conclusion similar to that in the LNC case.

\begin{figure}
\centering
\includegraphics[width=9.2cm]{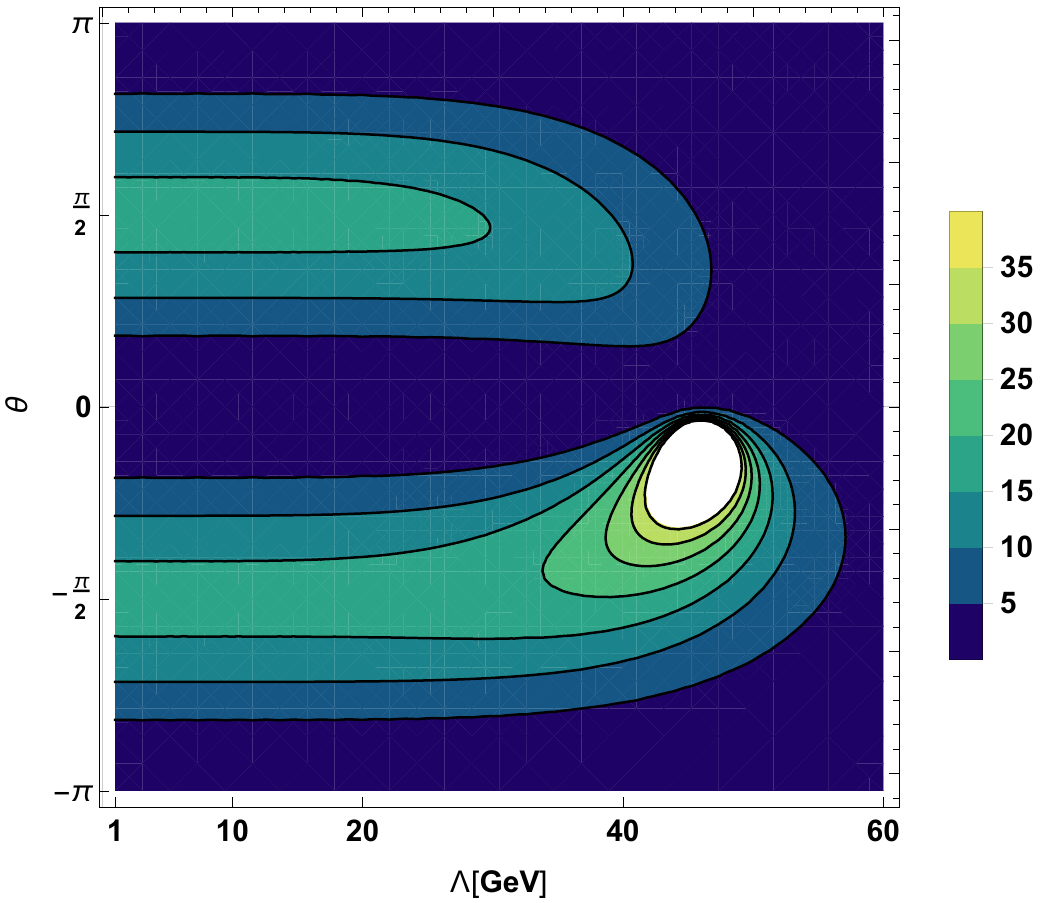}
\includegraphics[width=8cm]{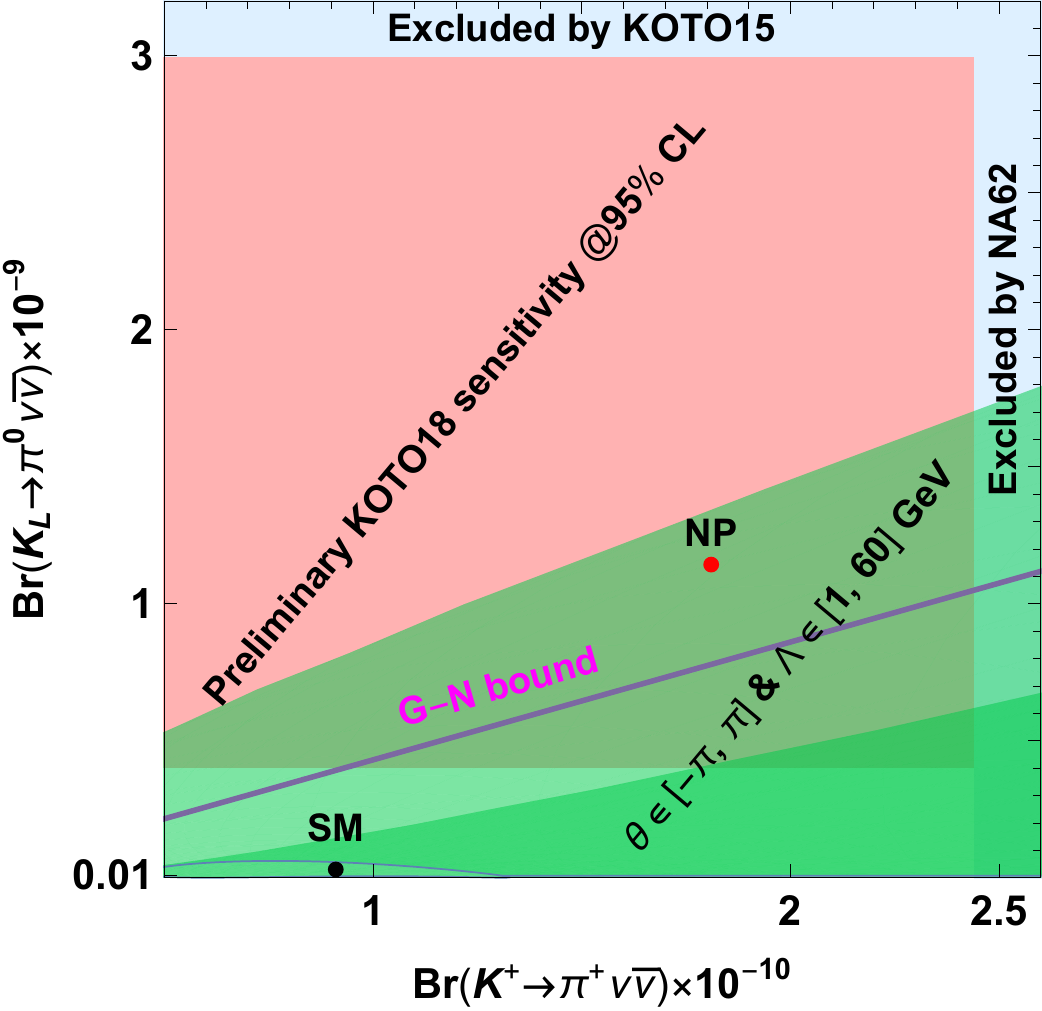}
\caption{Left panel: contours of $\mathcal{B}(K_L\to\pi^0\nu\bar{\nu})/\mathcal{B}(K^+\to\pi^+\nu\bar{\nu})$ values on the $\Lambda$-$\theta$ plane from the $\Delta I=1/2,3/2$ operators in the example NP model discussed in the text. Right panel: the branching-fraction predictions of the model compared to an interpretation~\cite{Kitahara:2019lws} of recent KOTO results~\cite{koto,Ahn:2018mvc} and to the latest NA62 limit~\cite{na62}, and the red dot labeled NP corresponds to $(\Lambda,\theta)=(39~\GeV, -\pi/4)$ in this model.} \label{fig1}
\end{figure}

To illustrate this general result, we consider a simple example involving the effective LNC Lagrangian ${\cal L}_{\rm int}=\hat c_a Q_a+\hat c_b Q_b+{\rm H.c.}$, where $Q_a$ and $Q_b$ are dimension-6 lepton-flavor-conserving operators which induce $\Delta I=1/2$ and $\Delta I=3/2$ transitions, respectively, and are given by
\begin{align}
Q_{a}=&~i\bigg(\pi^-\partial_\mu K^+-{1 \over \sqrt{2}}\, \pi^0\partial_\mu K^0\bigg) \overline{\nu_\alpha}\gamma^\mu\nu_\alpha \,, &
Q_{b}=&~i\bigg(\pi^-\partial_\mu K^++\sqrt{2}\,\pi^0\partial_\mu K^0\bigg) \overline{\nu_\alpha}\gamma^\mu\nu_\alpha\,,
\end{align}
and $\hat c_a$ and $\hat c_b$ are their coefficients.
The first operator, $Q_a$, is already mentioned in Eq.~\eqref{lnc:example1} and can arise in the SM, while $Q_b$ can proceed from this $\Delta I=3/2$ dimension-9 quark-level operator:
\begin{eqnarray}\label{lnc:dim9operator}
\calO_{\scriptsize\mbox{dim-9}}^{\textsc{lnc}} &=& \big[ \overline{d_L}s_R\left( \overline{u_R}\gamma_\mu u_R-\overline{d_R}\gamma_\mu d_R\right) + \overline{u_L}s_R\overline{d_R}\gamma_\mu u_R\big] \overline{\nu_\alpha}\gamma^\mu\nu_\alpha \,.
\end{eqnarray}
If we furthermore impose the SM gauge symmetry, ${\cal O}_{\scriptsize\mbox{dim-9}}^{\textsc{lnc}}$ can originate from, for instance, the dimension-10 operator $\big[\overline QHP_Rs\big( \overline u\gamma_\mu P_Ru-\overline d\gamma_\mu P_R d\big) + \overline Q\tilde{H}P_Rs\, \overline d\gamma_\mu P_Ru \big] \overline{L}\gamma^\mu P_L L $, where $Q$, $L$, and $H$ ($u$, $d$, and $s$) here are the SM quark, lepton, and Higgs doublets (quark singlets) under the SU(2)$_L$ group, and $\tilde H=i\tau_2H^*$ with $\tau_2$ being the second Pauli matrix.
From ${\cal L}_{\rm int}$ we derive
\begin{eqnarray}
\mathcal{A}_{K^+\to\pi^+\nu_\alpha\bar\nu_\alpha} &=& (\hat c_a+\hat c_b)\, \overline{u_\alpha}\slashed{k}P_Lv_\alpha \,, ~~~ ~~~~
\mathcal{A}_{K_L\to\pi^0\nu_\alpha\bar\nu_\alpha} \,=\, -i[{\rm Im}(\hat c_a-2\hat c_b)]\, \overline{u_\alpha}\slashed{k}P_Lv_\alpha \,, ~~~~~
\end{eqnarray}
which in view of Eq.~\eqref{lnc:ratioofbr} translate into
\begin{eqnarray} \label{BFR}
{\mathcal{B}(K_L\to\pi^0\nu\bar\nu) \over\mathcal{B}(K^+\to\pi^+\nu\bar\nu)} &=& \frac{\tau_{K_L}}{\tau_{K^+}} {{\rm Im}^2(\hat c_a-2\hat c_b) \over \big[ {\rm Re}^2 (\hat c_a+\hat c_b) + {\rm Im}^2 (\hat c_a+\hat c_b) \big]} \,.
\end{eqnarray}
We can see that this ratio can take any value from zero to infinity when $\hat c_a$ and $\hat c_b$ move in the complex plane. For definiteness, we take $\hat c_a$ to be the central value of the SM prediction~\cite{He:2018uey} and suppose that $\hat c_b$ stems from $\calO_{\scriptsize\mbox{dim-9}}^{\rm LNC}$ in Eq.~\eqref{lnc:dim9operator} and has the form $\hat c_b=F_KF_\pi B e^{i\theta}/\Lambda^5$ where $F_{K(\pi)}$ is the kaon (pion) decay constant~\cite{Tanabashi:2018oca}, $B=-\langle \bar qq\rangle/(3F_\pi^2)\approx2.8\,\GeV$ where $\langle \bar qq\rangle$ is the quark condensate which measures the strength of the chiral symmetry breaking effect, $\theta$ is some phase, and $\Lambda$ represents the scale of NP responsible for $\calO_{\scriptsize\mbox{dim-9}}^{\textsc{lnc}}$. In Fig.~\ref{fig1} we display on the left panel  the contour plot for the branching-fraction ratio in Eq.\,(\ref{BFR}) on the $\Lambda$-$\theta$ plane. We depict the predictions of this toy scenario for $\Lambda\in[1,60]$\,GeV and $\theta\in[-\pi,\pi]$ with the green region on the right panel and compare them to an interpretation~\cite{Kitahara:2019lws} of the recent KOTO results~\cite{koto,Ahn:2018mvc} as well as to the latest NA62 limit~\cite{na62}.
Clearly this model has parameter space which can explain the anomaly in the new preliminary KOTO data~\cite{koto} but the NP scale has to be of order tens of GeV, as the left plot indicates, this will be discussed in detail in our upcoming publication.
\section{Discussions and conclusions}

In this paper we demonstrate that the Grossman-Nir bound is always respected independent of the $CP$ property of the $\Delta I=1/2$ operators, the number of quarks involved, and whether or not the kaon decays conserve lepton number and that the bound is only the result of the $\Delta I=1/2$ nature of the relevant local operators together with the limits of the neutrino masslessness and the kaon state $K_{L(S)}=\big[K^0+(-)\bar{K}^0\big]/\sqrt{2}$. However, when $\Delta I=3/2$ operators are included, the GN bound could be violated. Those quark-level $\Delta I=3/2$ operators first appear at dimension nine. We take the SM $\Delta I=1/2$ operator and one $\Delta I=3/2$ operator in a toy scenario to illustrate how the GN bound is violated explicitly. We will present elsewhere a more detailed and systematic study of dim-9 quark-level operators that can violate the bound in the framework of SM effective field theory.

\acknowledgements

This work was supported in part by the MOST (Grant No. MOST 106-2112-M-002-003-MY3), and in part by the Australian Government through the Australian Research Council.


\end{document}